\documentclass[conference]{IEEEtran}
\IEEEoverridecommandlockouts
\usepackage{cite}
\usepackage{amsmath,amssymb,amsfonts}
\usepackage{algorithmic}
\usepackage{graphicx}
\usepackage{textcomp}
\usepackage{xcolor}
\usepackage{enumitem}
\usepackage{tabularx}
\usepackage{graphicx}
\usepackage{amsfonts}
\usepackage{amsmath}
\usepackage{breqn}
\usepackage{etoolbox}
\usepackage{color}
\usepackage{float}

\usepackage{hyperref}

\usepackage{comment}
\usepackage{csquotes}
\usepackage{amsmath,amssymb}

\def\BibTeX{{\rm B\kern-.05em{\sc i\kern-.025em b}\kern-.08em
    T\kern-.1667em\lower.7ex\hbox{E}\kern-.125emX}}
    
\begin{document}

\def\baselinestretch{0.95}

\makeatletter
\def\ps@IEEEtitlepagestyle{%
  \def\@oddfoot{\mycopyrightnotice}%
  \def\@evenfoot{}%
}
\def\mycopyrightnotice{%
  {\footnotesize \textcolor{red}{\begin{tabular}[t]{@{}l@{}} This paper has been accepted for publication by the 2023 IEEE International Conference on Communications (ICC 2023). © 2023 IEEE. Personal use \\ of this material is permitted. Permission from IEEE must be obtained for all other uses, in any current or future media, including reprinting/republishing \\ this material for advertising or promotional purposes, creating new collective works, for resale or redistribution to servers or lists, or reuse of any copyrighted \\ component of this work in other works.\end{tabular}}}
  \gdef\mycopyrightnotice{}
}

\title{Investigating the Characteristics and Performance of Augmented Reality Applications on Head-Mounted Displays: A Study of the Hololens Application Store
}

\author{
\IEEEauthorblockN{Pubudu Wijesooriya, Sheikh Muhammad Farjad, Nikolaos Stergiou
\IEEEauthorblockA{University of Nebraska at Omaha \\ 
\{pwijesooriya, sfarjad, nstergiou\}@unomaha.edu}}
\and
\IEEEauthorblockN{Spyridon Mastorakis}
\IEEEauthorblockA{University of Notre Dame \\
mastorakis@nd.edu}
}

\makeatletter
\patchcmd{\@maketitle}
  {\addvspace{0.5\baselineskip}\egroup}
  {\addvspace{-1.5\baselineskip}\egroup}
  {}
  {}
\makeatother


\maketitle

\begin{abstract}
Augmented Reality (AR) based on Head-Mounted Displays (HMDs) has gained significant traction over the recent years. Nevertheless, it remains unclear what AR HMD-based applications have been developed over the years and what their system performance is when they are run on HMDs. In this paper, we aim to shed light into this direction. Our study focuses on the applications available on the Microsoft Hololens application store given the wide use of the Hololens headset. Our study has two major parts: (i) we collect metadata about the applications available on the Microsoft Hololens application store to understand their characteristics (e.g., categories, pricing, permissions requested, hardware and software compatibility); and (ii) we interact with these applications while running on a Hololens 2 headset and collect data about systems-related metrics (e.g., memory and storage usage, time spent on CPU and GPU related operations) to investigate the systems performance of applications. Our study has resulted in several interesting findings, which we share with the research community.
\end{abstract}

\begin{IEEEkeywords}
Hololens, Augmented Reality, Measurements
\end{IEEEkeywords}

\section{Introduction}
Augmented Reality (AR) applications based on Head-Mounted Displays (HMDs) have gained significant traction over the recent years~\cite{kim2018revisiting}. A variety of AR HMD applications has been developed and several headsets have been released, such as the Microsoft Hololens~\cite{hololens} and the Magic Leap~\cite{magic-leap}. Despite this rapid growth, it remains unclear what kinds of applications have actually been developed over the years and what resources (e.g., computing, memory, storage) they occupy when they run on HMDs--in other words, what their system performance is when these applications run on HMDs. This information can provide insights into real-world AR application use cases and inform future design and development decisions related to both AR applications and HMDs. 

In line with this argument, in this paper, we conduct a study of the characteristics and system performance of AR HMD applications. Our study focuses on the applications available on the Microsoft Hololens application store, given that the Hololens headset has been widely used over the recent years. Our contribution in this paper is two-fold: 

\begin{itemize}[wide, labelwidth=!, labelindent=0pt]

\item We collect metadata about the applications available on the Hololens application store. This metadata offers visibility into the characteristics of AR applications that have been developed over the years, such as their categories, target audience, hardware and software compatibility, pricing, and required headset permissions. 

\item We interact with each application for a duration of five minutes and collect systems related data about the performance and the resources (e.g., computing, memory, storage) occupied by each application while running on a Hololens 2 headset. The conclusions we draw based on this data can provide useful guidance for AR application and HMD developers.

\end{itemize}

To the best of our knowledge, our paper is among the first attempts to understand the nature and systems performance/impact of AR HMD applications that have been developed so far by the community. We make the data we have collected publicly available to the research community through the following link: \url{https://drive.google.com/drive/folders/1A4815PtyC6CIm-TnS8BgoT1DqehDVaQf?usp=share\_link}.


\section{Background and Motivation}

\label{sec:back}
Microsoft released HoloLens in 2016. Hololens was in development for years under the umbrella of Project Baraboo. It is the first HMD that provides users with the functionality of an AR environment with the additional feature of interacting with the environment. The device is based on the Windows Mixed Reality (WMR) platform that is equipped with Windows 10 and 11 operating systems. WMR is responsible for rendering virtual objects in the real-world environment and enables users to interact with the 3D virtual overlays imposed by the WMR. 
Following the progress of HoloLens, Microsoft released its second version named ``HoloLens 2'' (second generation of Hololens). HoloLens 2 supports the development of gaming and entertainment applications but also provides support for enterprise market and commercial customers. With an improved gesture tracking system and an enhanced field of view, HoloLens 2 provides an enhanced quality of experience to users compared to its previous generation.

\subsection{HoloLens Vs. Conventional AR HMDs}

\subsubsection{Holographic Processing Unit}
The Holographic Processing Unit (HPU) is probably the most essential component of HoloLens. It is a custom multiprocessor that manages the task of processing the information flowing from on-board sensors onto the headset, including the time-of-flight depth sensor, the inertial measurement unit, and head-tracking cameras. HPU helps with the pre-calculation of sensor data, gesture recognition, the creation of maps of user surroundings, and the maintenance of virtual 3D object positions in the real world.

\subsubsection{Shared Augmented Environment}

HoloLens provides the functionality of a shared environment surrounding users. Different users can share the same augmented environment by pairing their headsets. The communication channel built among the shared devices enables users to utilize gesture tracking and other actions in order to modify the projected virtual environment. This functionality aids in developing apps for collaborative activities~\cite{Vidal-Balea} 

\subsubsection{Gesture Tracking}

Unlike conventional AR HMDs, HoloLens does not primarily depend upon optical markers to ascertain virtual object coordinates. Virtual touch interactions, hand ray pointers, air tapping, and gazing provide adaptable and versatile gesture movements. The immersed environment coupled with HPU enables HoloLens to identify gestures.

\subsection{Motivation}


The Hololens headset has been among the most popular and widely-used AR HMDs. Nevertheless, the device is still to a considerable extent proprietary and limited information is available regarding its internal operation. Considering also the growth of the number of AR applications that have been developed over the recent years to run on HMDs, in general, and on Hololens, in particular, it is still unclear: (i) \textbf{what AR applications have been been developed over the years}--for example, what the nature of these applications is, when these applications were released, what software and hardware support such applications require, and what permissions they require when they run on an HMD; and (ii) \textbf{what system resources (e.g., storage, memory, computing) these applications need/occupy when they run on an HMD}. 

Gaining an understanding of (i) and (ii) can provide useful insights not only to the Human-Computer Interaction and Extended Reality (XR) community, but also to the computer systems community. Such insights can ultimately highlight AR application use cases that have been popular over the years as well as use cases that have been overlooked, and inform future efforts in AR application research and development as well as in the development of software and hardware for HMDs.


\subsection{Primary Goal}

In this paper, our goal is to take a first step towards the motivation we discussed above by studying the AR applications available on the Microsoft application store. Our study consists of two major parts: (i) we first collect and analyze application metadata (Section~\ref{Metadata}); and (ii) we then interact with these applications to collect data related to their system performance and resource usage when they run on the Hololens headset. 


\section{Application Metadata}
\label{Metadata}

In this section, we present our study of HoloLens applications metadata, which provides a descriptive overview of each application. Our goal through this study is to gain visibility into the applications that have been developed over the years.

\subsection{Collection}

We collected metadata for 451 AR applications through the following two sources:

\begin{itemize}[wide, labelwidth=!, labelindent=0pt]

\item \textbf{Application Web Portal:} The primary source for collecting metadata was the Microsoft Web Portal. 
Although the HoloLens web portal provided metadata for most applications, there were some applications not listed on the website.

\item \textbf{Hololens Headset:} We collected the metadata for the applications that were not listed on the web portal by browsing the application store directly on the headset. 

\end{itemize}
 
\subsection{Description}

The collected metadata contains various attributes about the HoloLens applications. These attributes include the application name, category, rating, publisher, developer, file size of each application, release date, permissions, language, and software and hardware support required for each application. These attributes help us understand the nature of applications that AR developers have implemented since the release of the first generation of HoloLens.


\subsection{Analysis}


\subsubsection{Linguistic Study}

\begin{figure}[htbp]
\vspace{-0.2cm}
\centerline{\includegraphics[width=0.68\columnwidth]{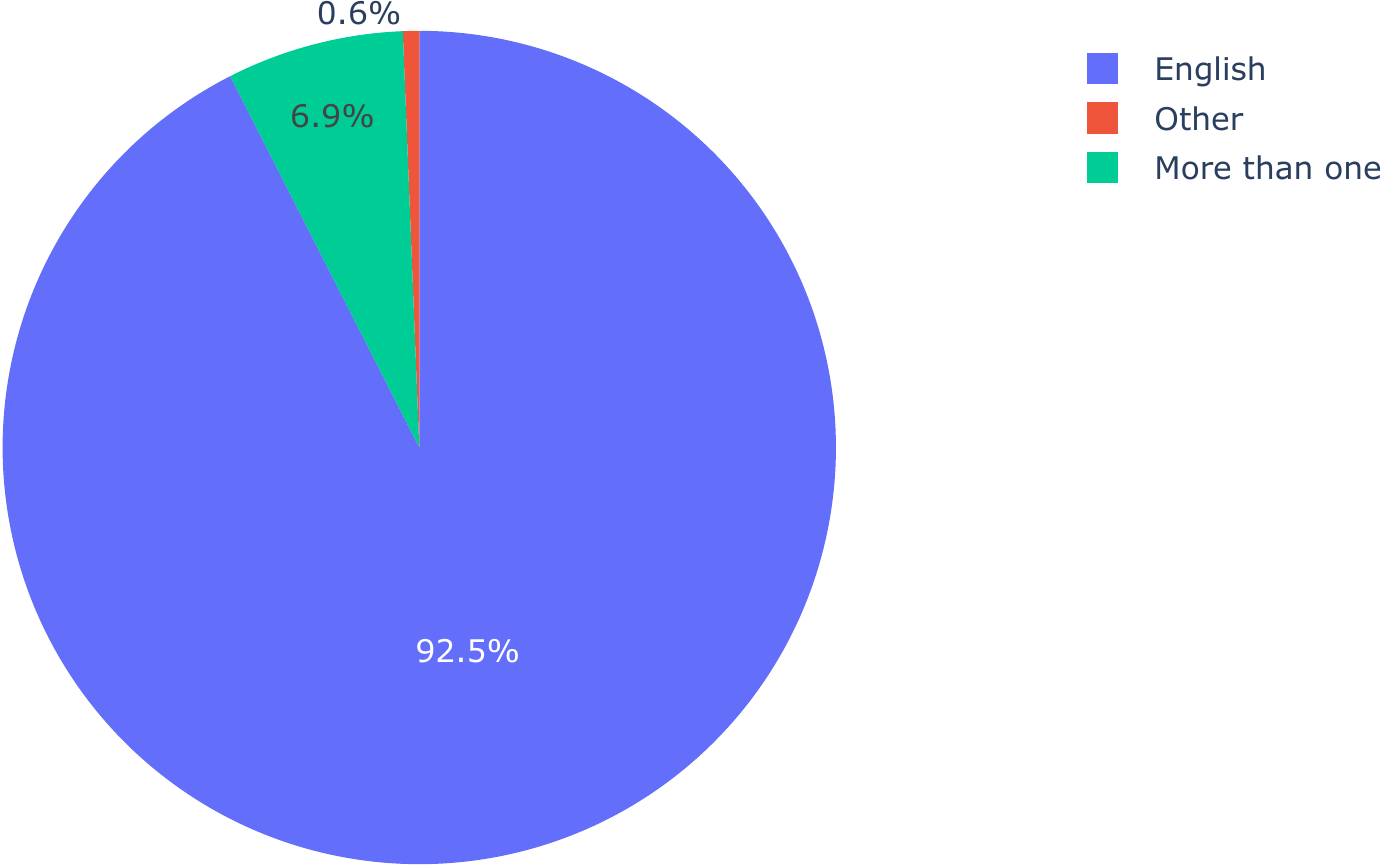}}
\vspace{-0.3cm}
\caption{A visualization of the languages supported by applications.}
\label{fig1}
\vspace{-0.3cm}
\end{figure}

There are 65 languages supported by Hololens applications. Our results, illustrated in Figure~\ref{fig1}, demonstrate that the language associated with the vast majority of applications is English. 6.9\% of the applications support more than one languages. For all these applications, one of the supported language is English, while other supported languages may include Spanish, German, French, Italian, Japanese, Chinese, and Portuguese. Finally, only three applications (0.6\% of all applications) exclusively support a language other than English (specifically German or Chinese).  


\subsubsection{Categorical Analysis}

In Table~\ref{tab:category}, we present the categories of the available Hololens applications. The majority of applications have been listed under entertainment, education, productivity, and business. Smaller portions of applications have been listed under several other categories, such as action \& adventure, utilities \& tools, simulation, puzzle \& trivia, developer tools, multimedia design, and sports. Finally, we have listed applications (3.33\% of the total available applications) under a category, which we call ``other''. This category aggregates less popular application categories, such as books \& references, health \& fitness, and racing \& flying.

\begin{table}[t]
\caption{Categories of Hololens applications.}
\vspace{-0.2cm}
\centering
\resizebox{0.5\columnwidth}{!}{%
\label{tab:category}
\begin{tabular}{|c|c|}
\hline
\textit{\textbf{Category}} & \textit{\textbf{\begin{tabular}[c]{@{}c@{}}Percentage of \\ applications (\%)\end{tabular}}} \\ \hline
Entertainment              & 14.86                                                                                        \\ \hline
Education                  & 14.63                                                                                        \\ \hline
Productivity               & 14.41                                                                                        \\ \hline
Business                   & 10.86                                                                                        \\ \hline
Action \& adventure        & 5.54                                                                                         \\ \hline
Utilities \& tools         & 4.21                                                                                         \\ \hline
Simulation                 & 3.99                                                                                         \\ 
\hline
Puzzle \& trivia           & 3.33                                                                                         \\
\hline
Other                      & 3.33                                                                                         \\ 
\hline
Developer tools            & 2.66                                                                                         \\ \hline
Multimedia design          & 2.44                                                                                         \\ \hline
Sports                     & 1.99                                                                                         \\ \hline
Strategy                   & 1.99                                                                                         \\ \hline
Family \& kids             & 1.99                                                                                         \\ \hline
Card \& board              & 1.55                                                                                         \\ \hline
Medical                    & 1.33                                                                                         \\ \hline
Music                      & 1.33                                                                                         \\ \hline
News \& weather            & 1.33                                                                                         \\ \hline
Photo \& video             & 1.33                                                                                         \\ \hline
Shooter                    & 1.11                                                                                         \\ \hline
Shopping                   & 1.11                                                                                         \\ \hline
Navigation \& maps         & 1.11                                                                                         \\ \hline
Travel                     & 1.11                                                                                         \\ \hline
Lifestyle                  & 1.11                                                                                         \\ \hline
\end{tabular}
}
\vspace{-0.3cm}
\end{table}



\subsubsection{Target Audience}

\begin{figure}[htbp]
\centerline{\includegraphics[width=0.68\columnwidth]{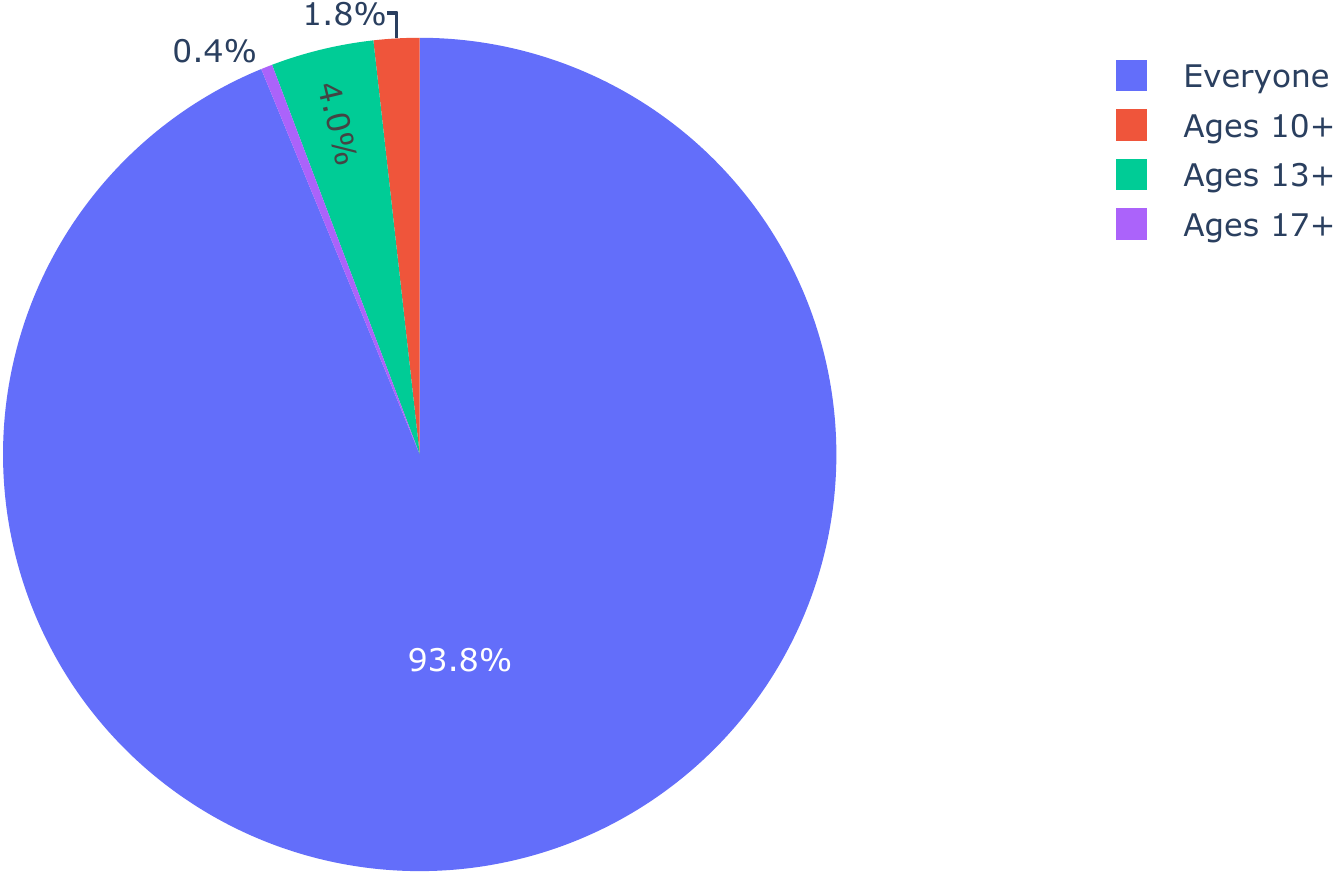}}
\vspace{-0.2cm}
\caption{The target audience of Hololens applications with respect to age.}
\label{fig3}
\vspace{-0.6cm}
\end{figure}

When considering the factor of age, as presented in Figure~\ref{fig3}, we observe that 93.8\% of the applications is suitable for everyone regardless of age. About 1.8\% and 4\% of the applications are exclusively for individuals who are at least 10 and 13 years old respectively. Two applications require users to be at least 17 years old. 

\subsubsection{Permissions Requested by Applications}

In Table~\ref{tab:permissions}, we present the permissions requested by Hololens applications. The majority of applications request access to current and past user surroundings, the microphone, and the Internet. A considerable portion of applications also requested that the headset acts as a server (typically used by applications to receive incoming data from the Internet as well as send logs for debugging purposes to application developers), access to the local network (typically used by applications that communicate across a local area network and/or share data among various local devices), and access to the webcam. Other permissions requested by applications include access to various libraries and files stored on the headset (e.g., pictures, videos, music), access to the current location of the user, access to gaze input (typically used by interactive applications), and access to 3D objects. 

An interesting observation based on the collected metadata is that a small portion of applications requested access to all system resources, the username and even the password of the user's account, as well as the permission to run as an administrator. Granting such permissions could result in security risks for users. For example, the user's Microsoft account credentials could be compromised if applications have access to them. The headset itself could be also compromised, if applications are allowed to run with administrator privileges.

\begin{table}[th]
\caption{Permissions requested by Hololens applications.}
\vspace{-0.2cm}
\centering
\label{tab:permissions}
\resizebox{0.7\columnwidth}{!}{%
\begin{tabular}{|c|c|}
\hline
\textit{\textbf{Permission}}                                             & \textit{\textbf{\begin{tabular}[c]{@{}c@{}}Percentage of \\ applications (\%)\end{tabular}}} \\ \hline
Access to current and past surroundings                                  & 76.50                             \\ \hline
Access to microphone                                                     & 73.60                             \\ \hline
Access to Internet connection                                            & 64.97                             \\ \hline
Access to Internet connection and act as a server                        & 40.58                             \\ \hline
Access to home or work network                                           & 36.58                             \\ \hline
Access to webcam                                                         & 32.59                             \\ \hline
Access to pictures' library                                              & 13.52                             \\ \hline
Access to video library                                                  & 8.20                              \\ \hline
\begin{tabular}[c]{@{}c@{}}Access to gaze input (typically used for \\ interactions within an application)\end{tabular}       & 7.54                              \\ \hline
Communicate with paired Bluetooth devices                                & 7.09                              \\ \hline
\begin{tabular}[c]{@{}c@{}}Other (e.g., screen projection, access to contacts, \\ run admin privileges, access to account's username \\ and password, access to device's certificates)\end{tabular}                                                                    & 7.09                              \\ \hline
\begin{tabular}[c]{@{}c@{}}Access to devices that support the Human Interface \\ Device (HID) protocol\end{tabular} & 6.65                              \\ \hline
Access to 3D Objects                                       & 5.76                              \\ \hline
Access to music library                                                  & 4.88                              \\ \hline
Access to current location                                               & 4.43                              \\ \hline
Use data stored on an external storage device                            & 3.77                              \\ \hline
No information                                                                  & 3.55                              \\ \hline
\begin{tabular}[c]{@{}c@{}}Access to devices that support Near \\ Field Communication (NFC) services\end{tabular}   & 3.32                              \\ \hline
Access to all system resources                                           & 2.66                              \\ \hline
Access to user account's username and picture                            & 2.44                              \\ \hline
\begin{tabular}[c]{@{}c@{}}Discover and launch applications \\ on other devices that user is signed into\end{tabular}       & 2.22                              \\ \hline
Access to device's voice over IP services                                & 1.55                              \\ \hline
Scan and connect to WiFi networks                                        & 1.55                              \\ \hline
Access to enterprise domain credentials                                  & 1.33                              \\ \hline
\begin{tabular}[c]{@{}c@{}}Access to the media player \\ while running in the background\end{tabular}               & 1.11                              \\ \hline
\end{tabular}
}
\vspace{-0.6cm}
\end{table}


\subsubsection{Supported Hardware Architectures}

\begin{figure}[htbp]
\centerline{\includegraphics[width=0.68\columnwidth]{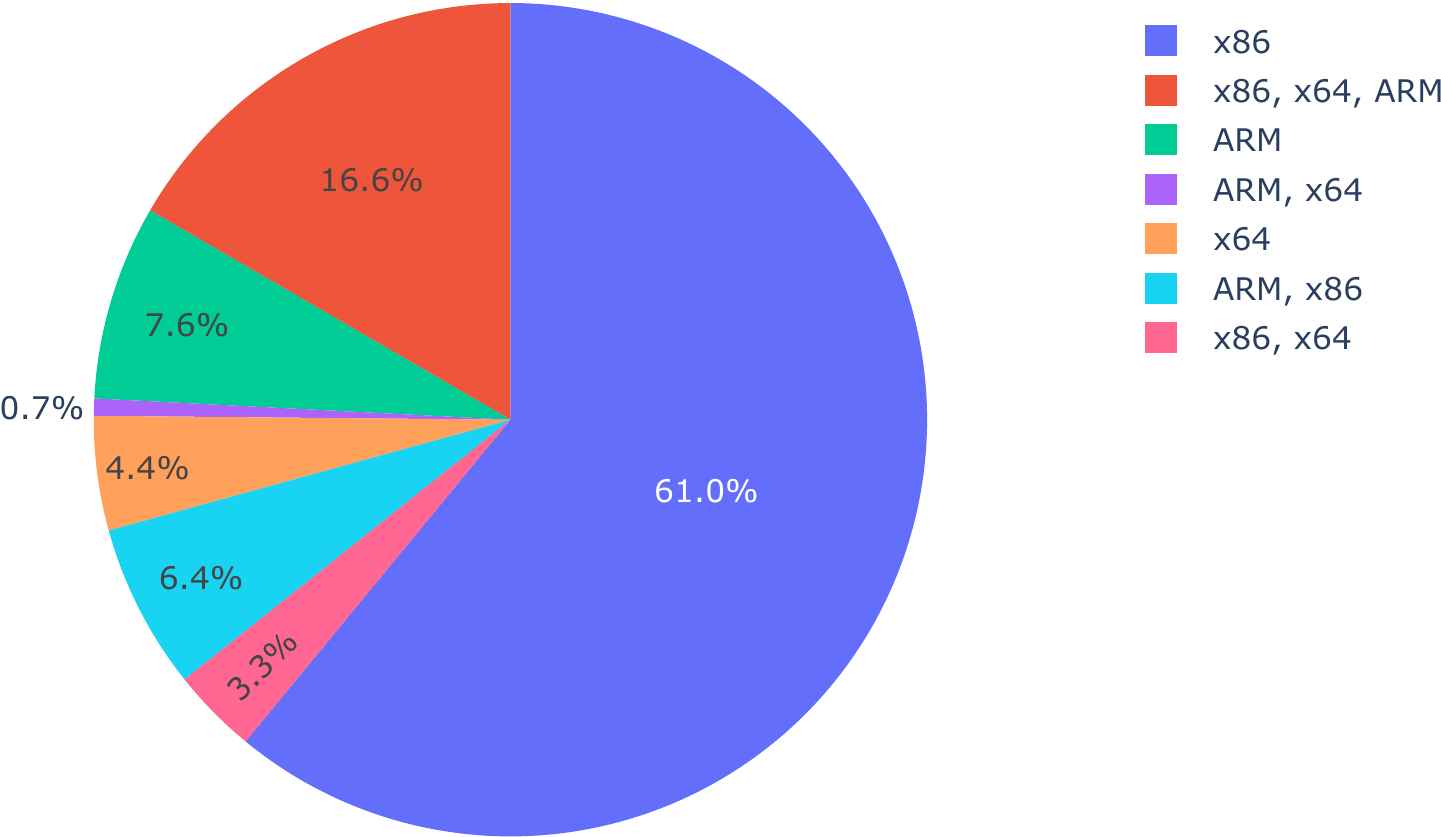}}
\vspace{-0.3cm}
\caption{The hardware architectures supported by Hololens applications.}
\label{fig5}
\vspace{-0.2cm}
\end{figure}

In Figure~\ref{fig5}, we present the hardware architectures supported by Hololens applications. The collected metadata indicate that the majority (61\%) of the available applications is compatible with only x86 architectures. This exclusive compatibility with x86 architectures restricts the operation of these applications only to the first generation of HoloLens (Hololens 1). One could argue that applications supporting the x86 architecture could be translated to support the ARM architecture of the second generation of Hololens (Hololens 2). This is true, however, the overall process is quite tedious and requires access to the application source code. 


As mentioned above, the second generation of HoloLens features an ARM architecture. About 7.6\% of the available applications support exclusively the ARM architecture. About 4.4\% of the available applications exclusively support x64 architectures. Our understanding is that these applications are either used to pair the headset with a computer/peripherals or are compatible with third-party headsets (e.g., Dell, Acer, and Samsung) that support the Windows Mixed Reality platform. 
16.6\% of the available applications support all three architectures (x86, x64, and ARM), while 6.4\% of applications support both x86 and ARM. Finally, smaller portions of the available applications support x86 and x64 as well as ARM and x64.

\subsubsection{Software Support}

\begin{figure}[htbp]
\vspace{-0.3cm}
\centerline{\includegraphics[width=0.68\columnwidth]{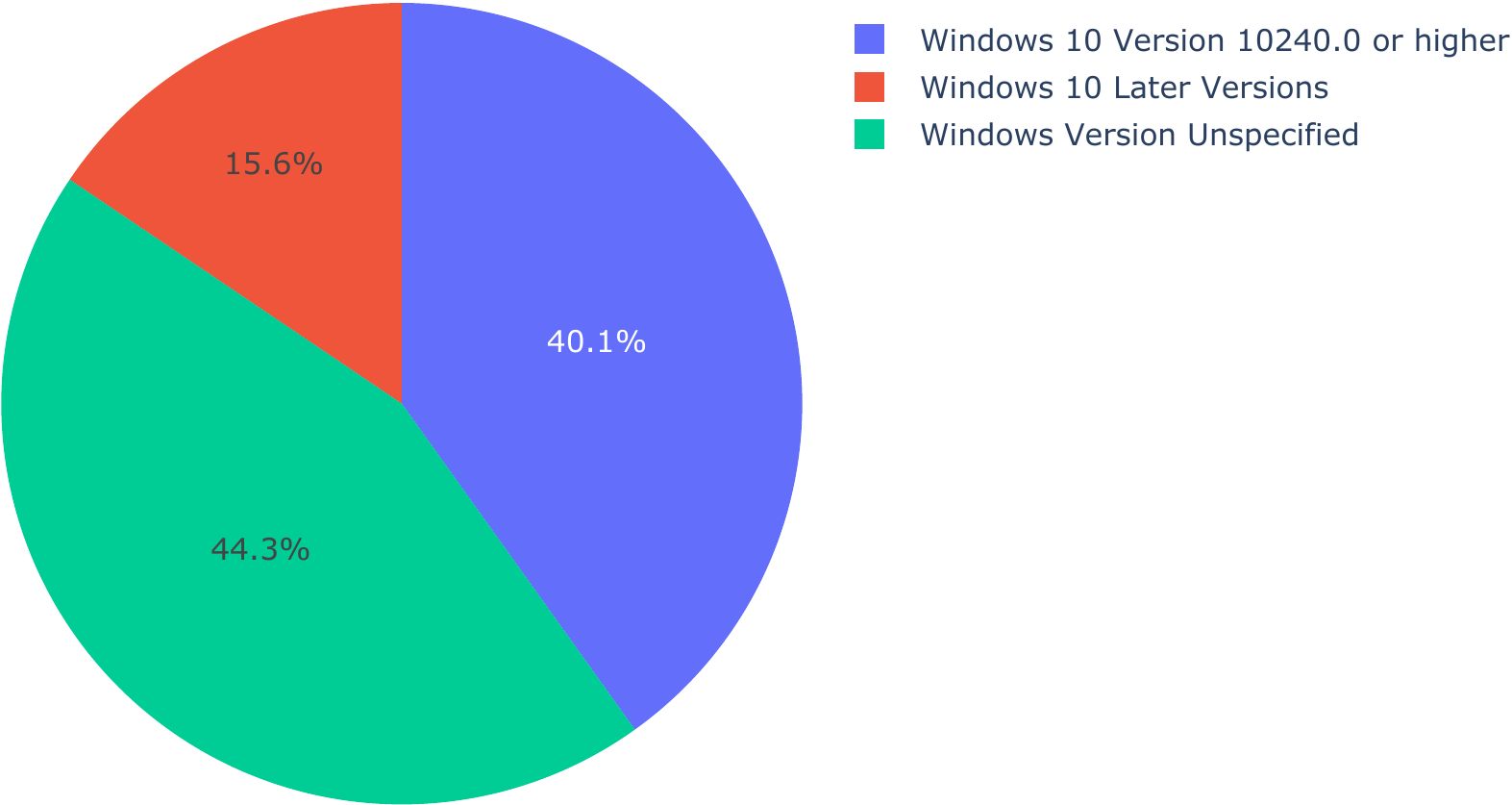}}
\vspace{-0.2cm}
\caption{The minimum software support required by Hololens applications.}
\label{fig6}
\vspace{-0.1cm}
\end{figure}

In Figure~\ref{fig6}, we present the minimum software support required by Hololens applications. This figure indicates that all applications require software support for, at least, Windows 10. The majority of applications (55.7\%) require a specific version (release) of Windows 10 or a newer version, while the rest (44.3\%) do not identify a minimum required Windows version. 40.1\% of the applications have as a minimum requirement Windows 10 build 10240. This is the original release of Windows 10, which became available to the general public on 29 July 2015. Finally, 15.6\% of the applications require a later version (release) of Windows 10. 

\subsubsection{Application File Size}

\begin{figure}[htbp]
\vspace{-0.3cm}
\centerline{\includegraphics[width=0.68\columnwidth]{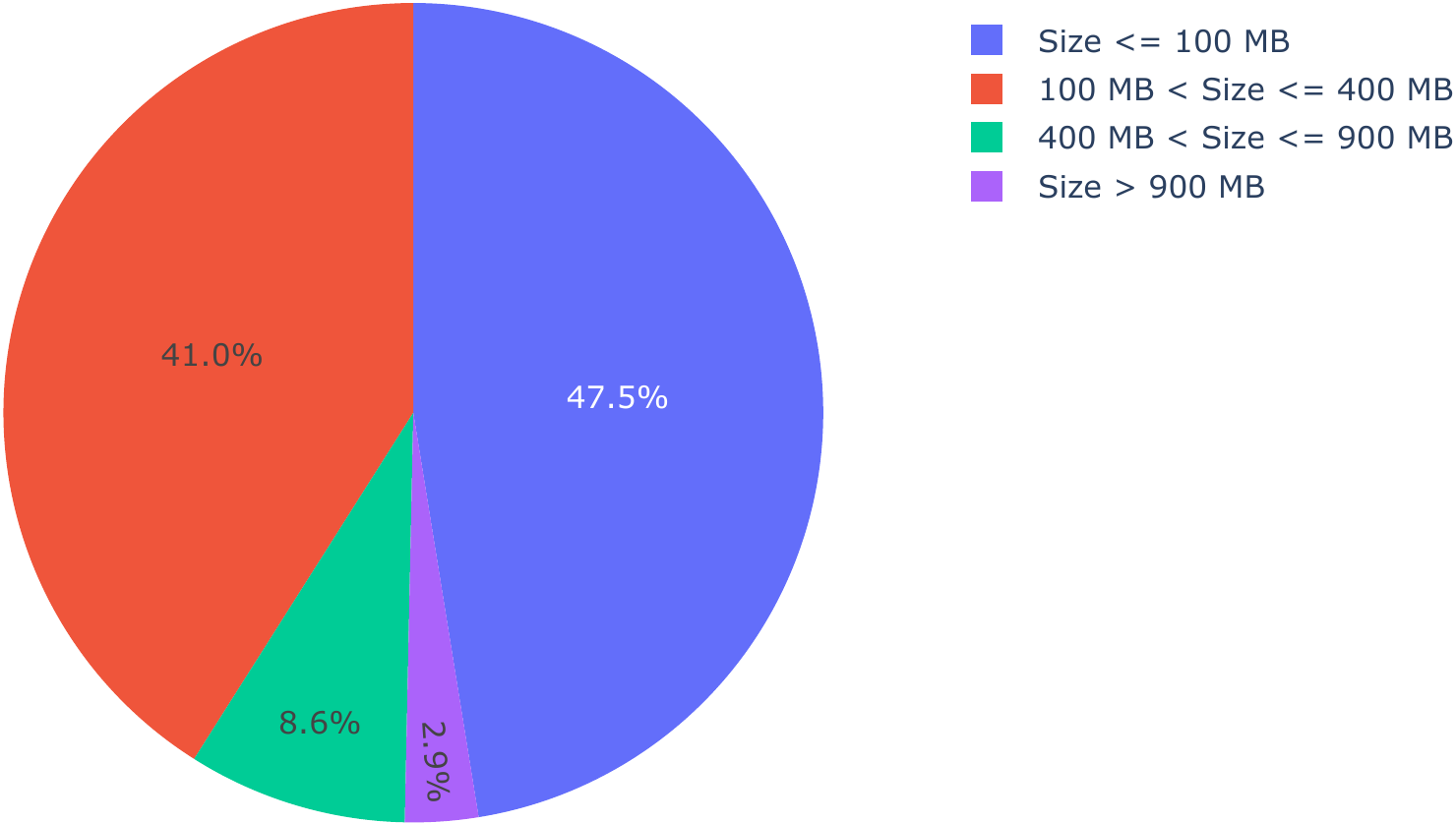}}
\vspace{-0.3cm}
\caption{The file sizes of applications as listed on the Hololens App Store.}
\label{fig7}
\vspace{-0.2cm}
\end{figure}

In Figure~\ref{fig7}, we present the application file sizes as listed on the Hololens App Store. 47.5\% of the available applications are smaller than 100 MB. 41\% of the applications have file sizes between 100 MB and 400 MB. Finally, 8.6\% of the available applications have file sizes between 400 MB and 900 MB, while only 2.9\% of the available applications are larger than 900 MB.

\subsubsection{Year of Release}

\begin{figure}[htbp]
\centerline{\includegraphics[width=0.68\columnwidth]{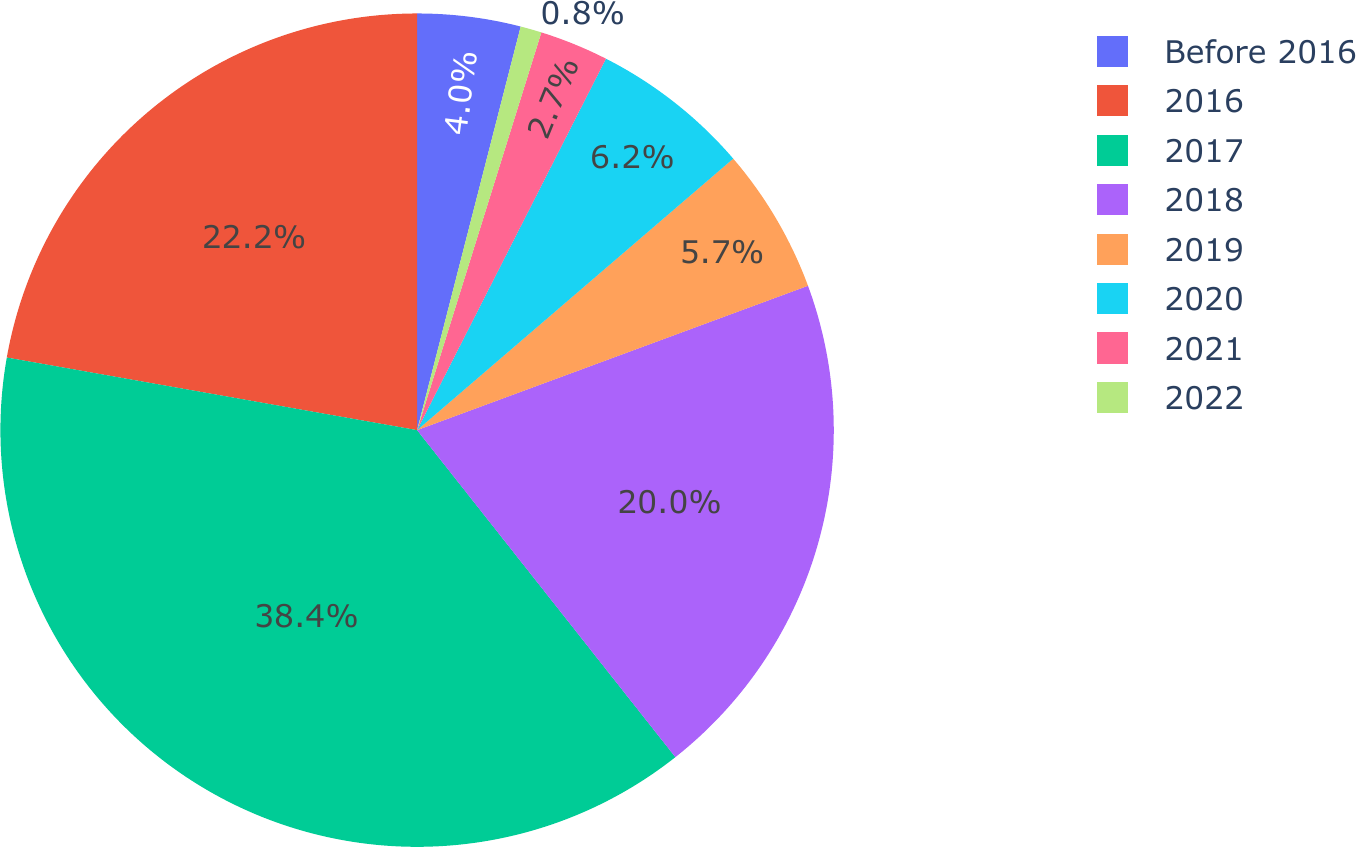}}
\vspace{-0.3cm}
\caption{A categorization of Hololens applications based on their release year.}
\label{fig8}
\vspace{-0.3cm}
\end{figure}

As shown in Figure~\ref{fig8}, the majority of applications were developed between 2016 and 2018. An interesting observation is that even before the official release of the first generation of Hololens in 2016, there were applications that had already been developed. Our understanding is that such applications were used during the development period of Hololens under the umbrella of Project Baraboo. Finally, the collected metadata indicate that only 15.4\% of the available applications have been released in 2019 and afterwards.


\subsubsection {Application Pricing}

\begin{figure}[htbp]
\vspace{-0.3cm}
\centerline{\includegraphics[width=0.68\columnwidth]{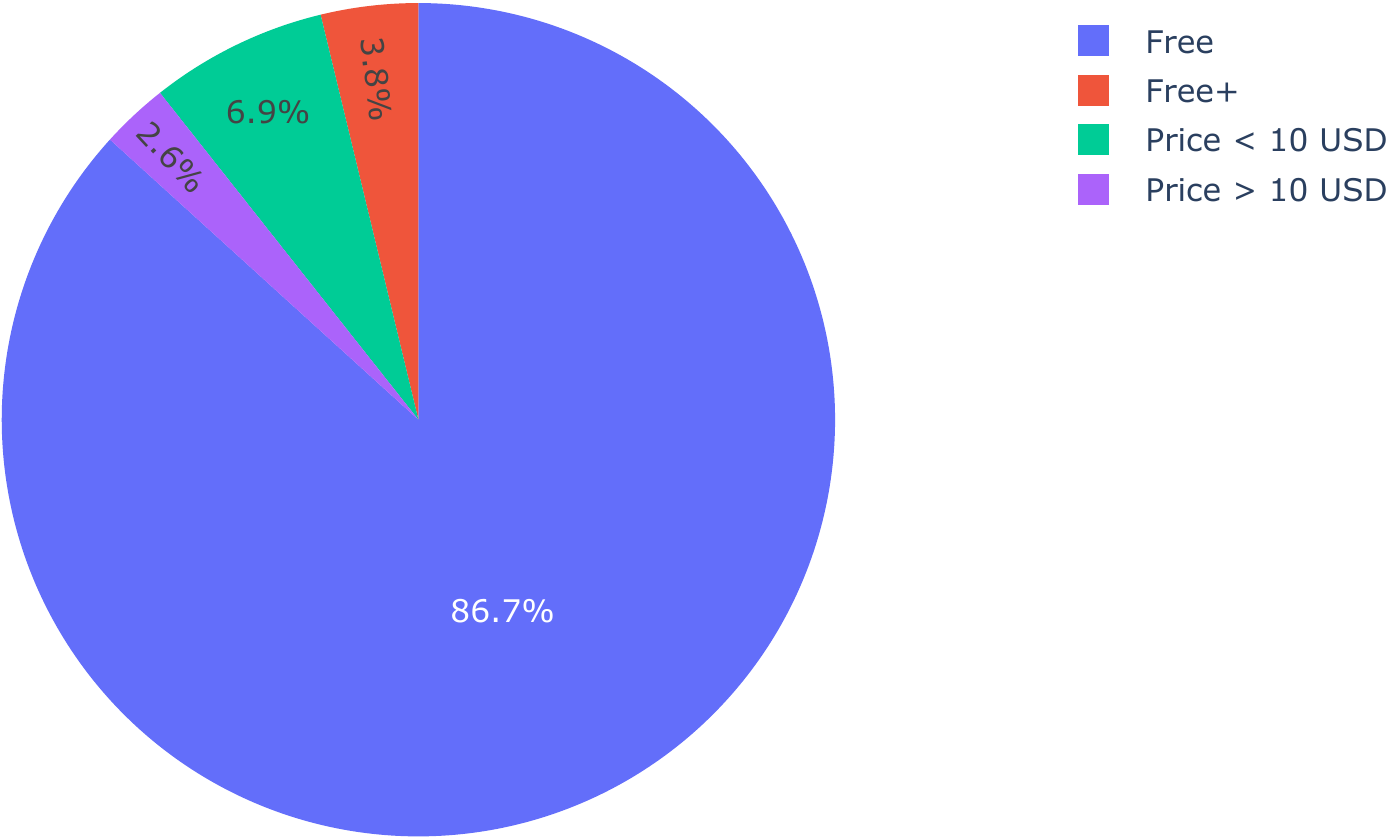}}
\vspace{-0.3cm}
\caption{A categorization of the pricing of Hololens applications.}
\label{fig9}
\vspace{-0.4cm}
\end{figure}

In Figure~\ref{fig9}, we present a categorization of the Hololens application pricing. The vast majority of applications (86.7\%) are free, while 3.8\% are the so called ``free+'' applications meaning that they are free for users to download, but optional in-application purchases (e.g., for access to advanced features) are available. The applications that require an upfront payment for users to download are about 9.5\% of all applications. Most of them (6.9\% of all applications) cost less than 10 USD, while 2.6\% of all applications cost more than 10 USD. The most expensive two applications available on the app store cost 999.99 USD each.

\section{Systems Related Data}          
\label{TraceFiles}

In this section, we collected trace files that demonstrate the systems performance of applications while they run on a Hololens 2 headset. We collected such traces for 141 applications compatible with the Hololens 2 ARM architecture.

\subsection{Collection}

We collected trace files for all applications compatible with a Hololens 2 headset while interacting with each application for a time duration of five minutes (300 seconds). To collect these traces, we used the Windows device portal, which is a web server that comes with Windows devices and provides tools to monitor the performance of a Windows device in real-time~\cite{windows-portal}. To capture data for various system-level parameters during the operation of the headset, we created custom performance recording profiles and configured the Windows Device Portal instance running on the headset with these profiles. We stored the captured data into Event Trace Log (ETL) files.


\subsection{Description}

We inspected the ETL files through the Windows performance analyzer and automated the extraction of data from these files 
through Python scripts. We focused on data related to the following categories of system resources: Central Processing Unit (CPU), Graphics Processing Unit (GPU), memory, and storage. Through the investigation of data related to these types of headset resources while applications are running on the headset, we can draw conclusions about the systems performance and the headset resources needed for the application operation.



\subsection{Analysis}


\subsubsection{CPU Resources}

In Figure~\ref{fig:time}, we present the time that the headset spent on performing CPU related operations while running the available applications. An interesting observation stemming from our results is that the CPU time for about 45\% of the applications exceeds the duration of our interaction with each of these applications, which was five minutes (300 seconds). This is due to the fact that the headset has eight CPU cores and, as a result, the applications use multiple CPU cores in parallel to increase performance. In this context, our results include the time spent on CPU related operations by all CPU cores cumulatively during the operation of applications.

In comparison to the time spent on GPU and storage related operations (which we discuss in detail in the rest of this section), our results demonstrate that the headset spent more time on CPU related operations. This is due to the fact that CPU is the resource used for most types of computations/operations needed by applications, such as processing input(s), animations, and physics. On the other hand, GPU is used for a limited set of operations that mostly include 3D object rendering, while storage is used even less frequently, since applications do not need substantial persistent data storage (i.e., data that persists after the headset is off).

Further analysis of the CPU related data we collected demonstrates that the headset mostly uses four out of the eight available CPU cores during the operation of applications. The other four CPU cores have a utilization that stays well under 1\% (typically close to 0.1\%) on average.

\subsubsection{GPU Resources}

In Figure~\ref{fig:time}, we present the time that is spent by the headset on performing GPU related operations while running the available applications. Our results indicate that the headset spent less time on GPU related operations than on CPU related operations. To our understanding, the main reason is that the headset's GPU is used for a limited (but still demanding) set of operations and mostly the rendering of 3D objects in comparison to the CPU, which is used for the execution of various other operations. 

Further analysis of our results demonstrates that the majority of applications (about 57\%) require the usage of more than 60\% of the headset's GPU resources while running on the headset. On the other hand, about 24\% of the available applications use less than 20\% of the headset's GPU resources during their operation. Applications that are ``GPU-hungry'' (i.e., require the use of most of the headset's available GPU resources) typically feature immersive graphics and involve extensive 3D object rendering operations, while applications that feature ``lighter'' graphics result in a lower GPU usage.

\subsubsection{Memory Resources}

In Figure~\ref{fig:memory}, we present the memory usage (consumption) of the Hololens headset while running the available applications. The total available memory on the headset is 4GB. Our results demonstrate that 6.5\% of the applications result in memory consumption of less than 3GB. 48.9\% of the applications result in memory consumption between 3GB and 3.5GB while running on the headset. 44.6\% of the applications result in more than 3.5GB of memory consumption while running on the headset. 

Further analysis of our results demonstrates that about 12\% of the available applications fully occupy the available memory resources at various moments during their operation. Further analysis also indicates that applications, which fully occupy the available memory resources during their operation, experience a substantial amount of page faults. Page faults happen when applications need to access data that should have been stored onto the main memory, however, such data had to be stored onto the headset's storage resources because the main memory was fully occupied. In such cases, it takes more time for applications to access such data, since access to storage resources is, in general, slower than access to the main memory as we further discuss in Section~\ref{subsec:storage} below.


\subsubsection{Storage Resources}
\label{subsec:storage}

In Figure~\ref{fig:memory}, we present the storage usage (consumption) by Hololens applications while running on the headset. Our results demonstrate that the storage usage of applications is limited, since the vast majority of applications occupy less than 0.5 GB of storage. In comparison to the memory usage results discussed in the previous subsection, our results indicate that applications occupy more memory than storage. Our interpretation is that this is happening because access to the main memory is faster than storage. In addition, storage, in general, is used when the main memory becomes fully occupied (memory paging) as well as to store data that needs to persist once the headset is turned off. However, most of the data stored by applications does not need to persist once the headset is turned off, since such data is mostly used during the operation of applications.

\begin{figure}[htbp]
\vspace{-0.4cm}
\centerline{\includegraphics[width=0.65\columnwidth]{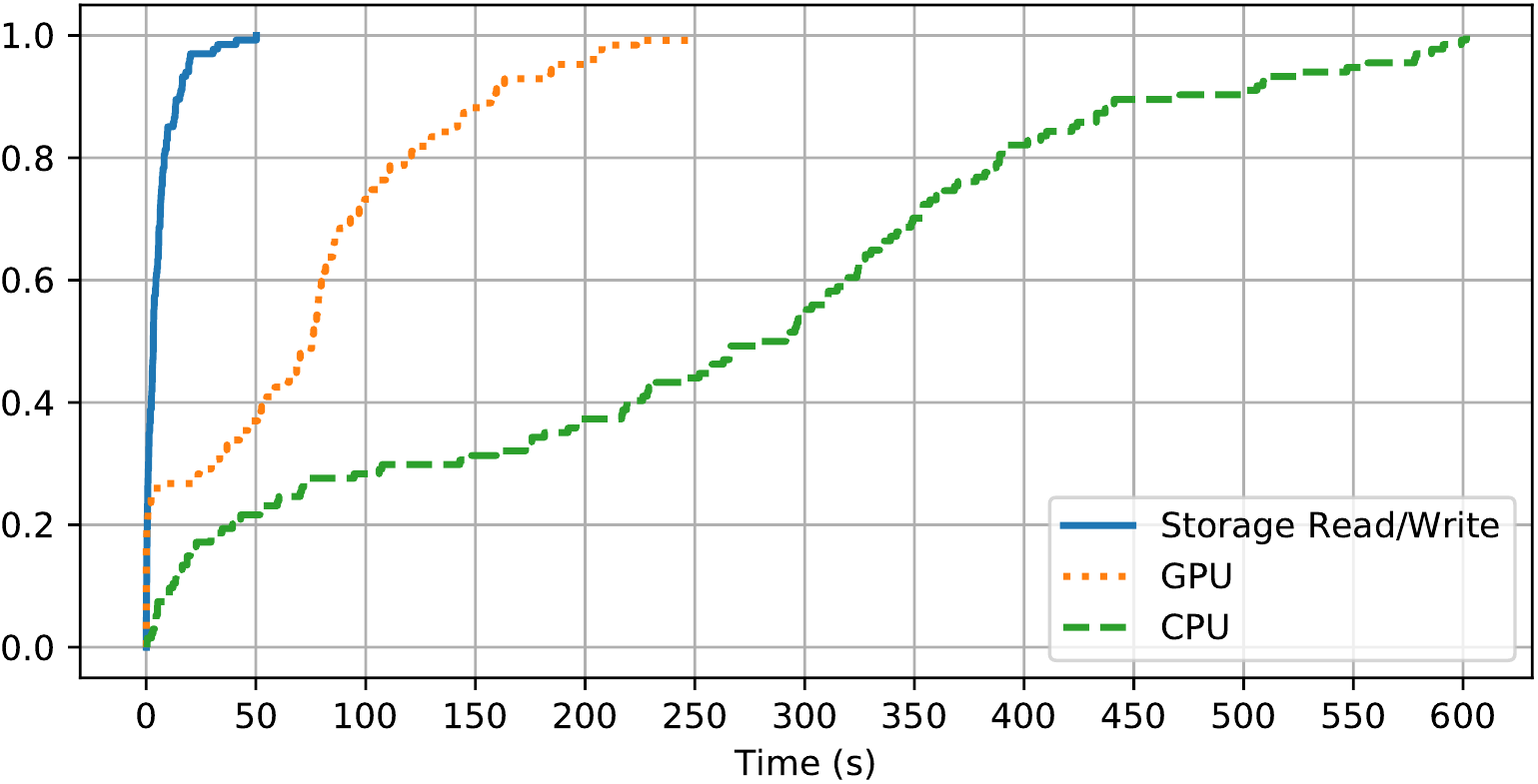}}
\vspace{-0.35cm}
\caption{The CDF plots of the time spent by the Hololens headset on CPU, GPU, and storage related operations while running various applications.}
\label{fig:time}
\vspace{-0.4cm}
\end{figure}

\begin{figure}[htbp]
\vspace{-0.3cm}
\centerline{\includegraphics[width=0.65\columnwidth]{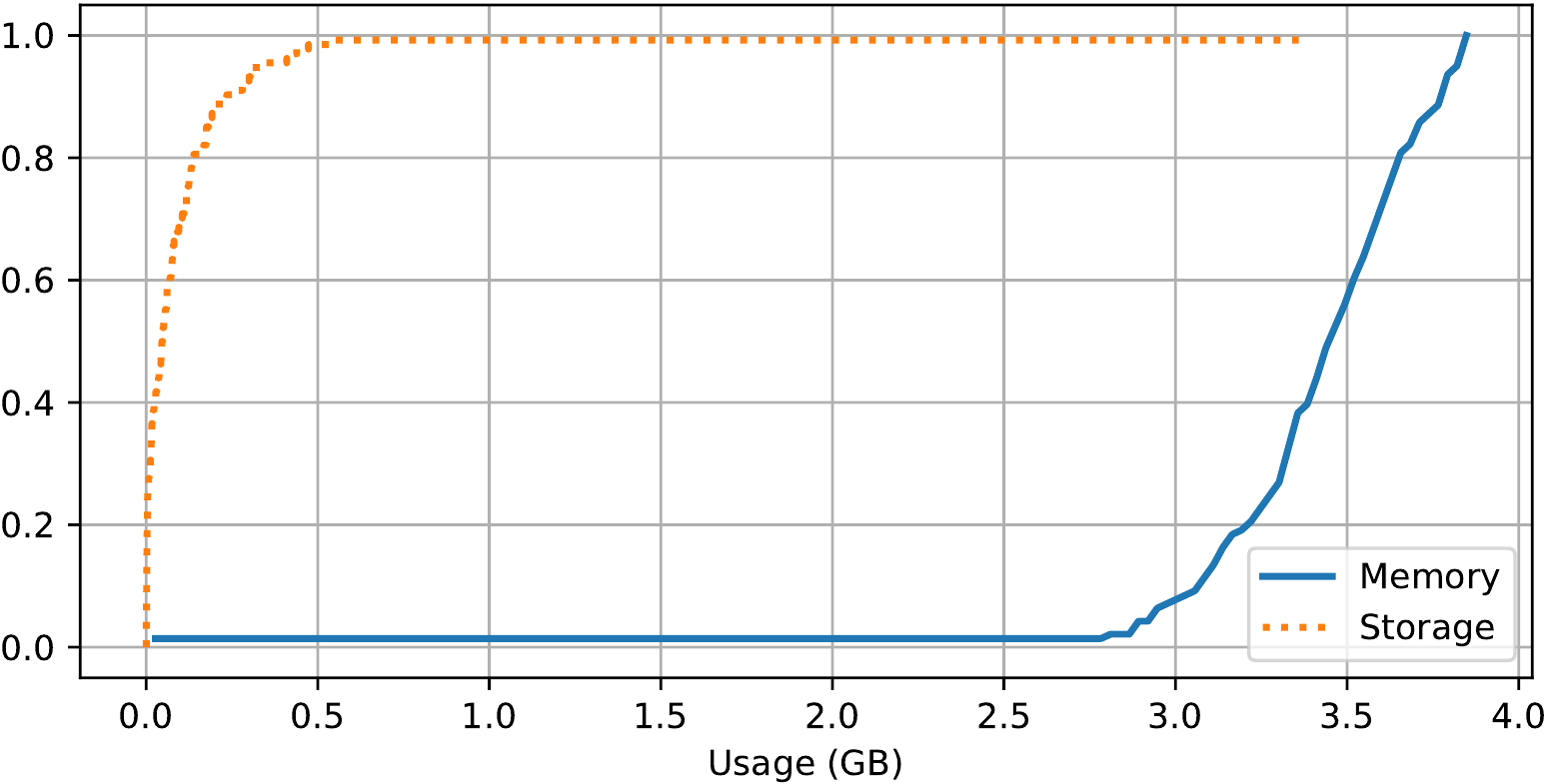}}
\vspace{-0.35cm}
\caption{The CDF plots of memory and storage consumption of the Hololens headset while running various applications.}
\label{fig:memory}
\vspace{-0.5cm}
\end{figure}

In Figure~\ref{fig:time}, we present the time that the headset spent on reading and writing data from/to its storage resources. In line with the results above that indicate limited usage of storage resources by applications, the usage of storage is low and faster than CPU/GPU operations. 

\section{Related Work}
\label{sec:related-work}

\textbf{Virtual Reality (VR) and AR use cases:} 
The community has utilized VR to create immersive environments for various application scenarios. Visualization and immersion help users interact and understand their surroundings in an effective way \cite{Erica}. Ritter et al. studied the response of karate athletes in a virtual environment~\cite{Ritter}. 
Iyer et al. proposed approaches for VR to enhance sports public relations and immersive journalism~\cite{Iyer}. 
The military and educational sectors used VR headsets for training, teaching, and mental health improvement \cite{Fan, Bell}. 
VR can also benefit the tourism industry \cite{Beck}. 
 
AR applications explore use cases where the users' perception of the world is augmented rather than creating a separate virtual world. For example, in manufacturing, workers can view instructions about how to assembly manufactured parts while working on the parts~\cite{nee2012augmented}. In education, students can interact with each other as well as teachers in an AR classroom setting, manipulating 3D objects to explore and understand different concepts~\cite{antonioli2014augmented}. In tourism, visitors can receive real-time feedback about famous sights while visiting these sights~\cite{yovcheva2012smartphone}. In clinical settings, medical personnel can use AR to provide effective clinical care~\cite{hilty2020review}. 





\noindent\textbf{Application benchmarking and auditing:} Application benchmarking and auditing are techniques to quantify application performance and identify security and privacy concerns respectively. Richoz et al. \cite{richoz2019benchmarking} benchmarked applications through classifiers running on mobile devices. Trimananda et al. \cite{Trimananda} presented a study of 140 Oculus VR applications, where they captured the network traffic generated by these applications and presented an analysis of the applications' privacy policies. Adams et al. \cite{Adams} explored the security and privacy aspects of VR through interviews with VR users and developers. 
Finally, Miller et al. \cite{Miller} studied the body motions of users in a VR environment and presented how they can aid the identification of personal characteristics.

\noindent\textbf{Other relevant work: }
Park et al. \cite{Park} conducted a review of 44 papers on the applications of HoloLens published between 2016 and 2020. They found that these applications can be divided into five categories: medical and surgical aids, medical education and simulation, industrial, architectural and civil engineering, and other engineering categories. 

\vspace{-0.1cm}

\section{Conclusion}
\label{sec:conclusion}

\vspace{-0.1cm}

In this paper, we presented a study of the AR applications available on the Microsoft Hololens application store. Based on our analysis, we drew the following four major conclusions:

\begin{itemize}[wide, labelwidth=!, labelindent=0pt]

\item The available AR applications have primarily focused on four categories: entertainment, education, productivity, and business. Applications have been developed in several other categories, but to a lesser extent.

\item Hololens applications typically request permissions, such as access to a user's past and present surroundings, a headset's microphone, and access to the local network for the headset to communicate through firewalls and share data with local devices. Such permissions could raise security and privacy concerns. In addition, a smaller portion of applications could result in much more serious security implications, since they request access to all system resources, access to user account credentials, and permissions to run as an administrator.

\item Most of the AR applications available for Hololens are compatible with the first generation of Hololens, thus being outdated and cumbersome to use today. This is due to the fact that not only the first generation of the Hololens headset has been discontinued, but also the software support provided for this headset is rather limited at this point (if existent at all).

\item CPU and memory are the resources predominantly occupied by AR applications while running on a Hololens 2 headset. Resources, such as GPU and storage, are used to a lesser extent during the operation of the applications. 

\end {itemize}

We ultimately believe that to make the use of XR pervasive in our everyday lives, an effort across disciplines, including Human Computer Interaction, computer systems and networking, and security and privacy, is needed. We hope that this paper will provide useful insights to researchers from these disciplines, contributing to the realization of this direction.




\section*{Acknowledgements}

This work is partially supported by the National Science Foundation  (awards CNS-2104700, CNS-2306685, CNS-2016714, and CBET-2124918), ACM SIGMOBILE, and the National Institutes of Health (award NIGMS/P20GM109090).

\bibliographystyle{unsrt}
\bibliography{References}

\end{document}